# Of Protein Size and Genomes

NEREIDE S. SANTOS-MAGALHÃES, HÉLIO M. DE OLIVEIRA
Federal University of Pernambuco
Department of Biochemistry, Laboratory of Immunopathology Keizo-Asami −LIKA
Signal Processing Group
Av. Prof. Moraes Rego 1235, 50670-901, Recife BRAZIL

E-mail: {nssm,hmo}@ufpe.br

*Abstract:* - An approach for approximately calculating the number of genes in a genome is presented, which takes into account the average protein length expected for the species. A number of virus, bacterial and eukaryotic genomes are scrutinized. Genome figures are presented, which support the average protein size of a species as a criterion for assessing life complexity. The human gene distribution in the 23 chromosomes is investigated emphasizing the genomic rate, the mean 'exon' length, and the mean 'exons' per gene. It is shown that storing all genes of a single human definitely requires less than 12 MB.

## 1 Introduction

The analysis of genomic information [1-2] is receiving wide attention, especially because of its importance in the early diagnosis of diseases, and novel tools are expected to emerge [3]. In living organisms, an assessment of their number of genes can be made before sequencing the complete genome. This can usually be derived taking into consideration some expected coding density. The probing starts with bacterial genomes; the evaluation of such genomes showed that the number of genes is, as a rule of thumb, numerically equal to the genome size expressed in kbp. An investigation of the average size of bacterial proteins reveals 350 amino acid residues as typical. Continuing with a much more intricate organism, the *C. elegans* was chosen, which has a genome of 99 Mbp and a genomic rate of 25%. Its protein size distribution has an average polypeptide length of 469 amino acids. A number of human proteins are quite long; serum albumin has 609 amino acid residues, collagen about 1,000, apolipoprotein B 4,536, and human Titin 26,926. It is therefore possible to predict an average human protein size at least as long as 600 amino acid residues. The aim of this paper is to show that the number *g* of genes of a genome can be estimated using the genome length *C* (bp), the genomic rate *R* and the average protein size $\overline{L}$ (expressed by the number of amino acid residues). The human genome will particularly be focused in the light of this approach.

## 2 Genes, Coding Density and Genomic Information

A DNA code is specified by the triplet DNA($C,R,\delta$), where *C* is genome size (bp), *R* is genomic rate and $\delta$ is coding density (genes/bp). *R* is defined here as the ratio between the number of protein-coding base pairs and the total number *C* of base pairs of the genome. This figure provides a clue to the redundancy of the code [4]. Further DNA coding parameters are *g*, $\overline{E}$ and *e*, where *g* is the number of genes of the genome, $\overline{E}$ is the average length of 'exons' and *e* is the average number of 'exons' per gene. The following relationships among parameters hold:

$$g = C/\delta \quad \text{and} \quad e.\overline{E} = R\delta. \qquad (1)$$

Furthermore, the amount of information on the genome can be estimated by considering 2 bits per coding base pair (for the sake of simplicity, Shannon information is not adopted here).

This note introduces an approach for deriving novel estimates of DNA code parameters, taking into account the average length of polypeptide chains of proteins expressed by the genes. For a given genomic rate *R*, the number of genes can be computed by

$$g = \frac{CR}{3\overline{L}} \quad \text{genes,} \qquad (2)$$

where $\bar{L}$ is the average number of amino acid residues (aa) of proteins. The coding density can also be estimated in terms of the expected protein size according to

$$\delta = \frac{C}{g} = \frac{3\bar{L}}{R} \text{ bp/gene.} \qquad (3)$$

For instance, the average bacterial protein is often around 300 amino acids long, and the genomic rate is typically in the range from 0.8 to 0.9. Bacteria usually have a coding density $\delta \approx 1,000$ bp/gene so their number of genes is, roughly speaking, numerically equal to the genome length expressed in kbp, i.e. $g \approx C/1,000$ (this is striking confirmed at http://www.cbs.dtu.dk/services/GenomeAtlas/).

Synthesized proteins from RNA translation have different lengths, usually ranging from 30 to more than 20,000 amino acids. An analysis of the protein length distribution in several microorganisms was previously reported [5].

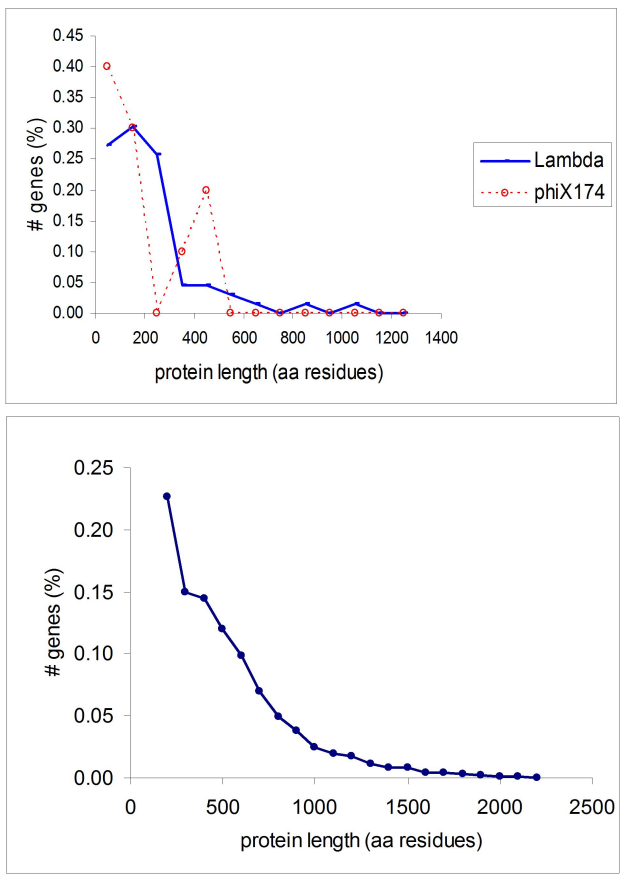

Fig. 1. Distribution of protein length for some simple organisms: (A) ΦX174 and Bacteriophage λ; and (B) *C. elegans*.

The mean of the protein size distribution depends on the complexity of the organism. Fig. 1 shows the protein size histograms for three straightforward organisms, the ΦX174 and the phage λ viruses (Fig.1A) and the *C. elegans* (Fig.1B).

The coding density of different chromosomes of lower eukaryotic species is roughly the same, i.e. shows only slight fluctuations from one chromosome to another in the same organism. The *S. cerevisiae* (*C*=12,057,849 bp, *g*=6,268 genes) has an average coding density 1,947 bp/gene considering its 15 chromosomes. The six chromosomes of the *C. elegans* (*C*=98,971,533 bp, *g*=17,585 genes) present an average coding density of 5,731 bp/gene. The coefficient of variation (CV %) of the coding density is 5.06 % for *S. cerevisiae,* and 1.72 % for the *C. elegans*. A reasonable evaluation for the coding density can therefore be used to derive a sound guesstimate of the number of genes. In contrast, higher eukaryotic cells are actually much more intricate: their coding density may fluctuate considerably from chromosome to chromosome. As a consequence, the task of estimating the number of genes becomes tricky. This piece of evidence is addressed in the next section.

Table 1. Eukaryotic coding density in chromosomes: *S. cerevisiae* (*C*=12,057,849 bp, *g*=6,268 genes); *C. elegans* (*C*=98,971,533 bp, *g*=17,585 genes). The coding density barely varies from one chromosome to another.

|      | *S. cerevisiae* |       |       | *C. elegans* |       |
|------|------|-------|-------|------|-------|
| Chr1 | 2,093 | Chr9  | 1,864 | ChrI   | 5,072 |
| Chr2 | 1,918 | Chr10 | 1,906 | ChrII  | 5,592 |
| Chr3 | 1,855 | Chr11 | 1,960 | ChrIII | 5,771 |
| Chr4 | 1,870 | Chr12 | 1,989 | ChrIV  | 6,312 |
| Chr5 | 2,090 | Chr13 | 1,841 | ChrV   | 4,899 |
| Chr6 | 2,144 | Chr14 | 1,854 | Chr X  | 6,740 |
| Chr7 | 1,891 | Chr15 | 1,908 |        |       |
| Chr8 | 2,017 | average | 1,947 bp/gene | average | 5,731 bp/gene |

(from http://www.cbs.dtu.dk/services/GenomeAtlas

Table 2 shows DNA parameters for some well-known genomes, which comprise the virus Φ*X174* [6], the microbial *M. genitalium* [7], *H. pylori* [8], *H. influenzae* [9], *S. aureus* [10], *B. subtilis* [11], *M. tuberculosis* [12], *E. coli* [13] and *X. fastidiosa* [14].

The average protein length was computed from Eqn(2). Clearly, an increase in the complexity of the organism is followed by a corresponding increase in the average protein length, as expected. According to Eqn(2), the average protein of *C. elegans* is 469 amino acids long, which is in agreement with its protein distribution [15]. The bottom four lines of Table 1 also include two possible scenarios for the human genome. The question mark suggests that the old-accepted estimate clearly underestimates the expected length of the average human protein.

An early and unsuccessful attempt to explain the complexity of living beings was the genome length. The so-called C-value paradox rapidly proved that this hypothesis was incorrect [16, 35]. The number of genes was afterward often supposed to be related to complexity. This reasoning partially biased people to expect more genes than human actually have. After all, how could one explain the fact that humans are so much more intricate than *Drosophila*? But is the *C. elegans* more complex than the *Drosophila*? It appears that life complexity is enormously sensitive to the protein length distribution. A potential measure that correlated with the complexity of beings could be its average protein size.

The genomic information gives an inkling of the file length required for storing only the protein-coding genes of the genome, without data compression. It is worthy of note that all genes of a bacterial genome can easily be stored in a single floppy disk (1.54 MB). Surprisingly, storing all genes of a single human will definitely require less than 12 MB (a typical CD has 700 MB and a PenDrive 256 MB available), albeit the entire the human DNA sequence requires about 1 GB.

Table 2. Features of a few sequenced genomes, emphasizing redundancy-related parameters (approximate values, source http://www.cbs.dtu.dk/services/GenomeAtlas/).

| Organism | genome size $C$ (Mbp) | coding density $\delta$ (bp/gene) | number of genes $g$ | genomic rate $R$ | average protein length | genomic information (Mbits) | redundancy $1-R$ (%) |
|---|---|---|---|---|---|---|---|
| $\Phi X174$ | 0.0054 | 538 | 10 | 1.00 | 180 | 0.01 | ~0 |
| λ bacteriophage | 0.0485 | 683 | 71 | 0.95 | 216 | 0.09 | 5 |
| *M. genitalium* | 0.58 | 1,208 | 480 | 0.90 | 363 | 1.04 | 10 |
| *H. pylori* | 1.67 | 1,066 | 1,566 | 0.89 | 316 | 2.97 | 11 |
| *H. influenzae* | 1.83 | 1,071 | 1,709 | 0.86 | 307 | 3.15 | 14 |
| *S. aureus* | 2.80 | 1,069 | 2,619 | 0.84 | 299 | 4.70 | 16 |
| *B. subtilis* | 4.21 | 1,025 | 4,106 | 0.87 | 297 | 7.32 | 13 |
| *M. tuberculosis* | 4.41 | 1,126 | 3,918 | 0.97 | 364 | 8.56 | 3 |
| *E. coli* | 4.64 | 1,082 | 4,289 | 0.87 | 314 | 8.08 | 13 |
| *X. fastidiosa* | 2.52 | 1,238 | 2,034 | 0.78 | 322 | 3.93 | 22 |
| *S. cerevisiae* | 12.06 | 1,924 | 6,268 | 0.70 | 450 | 17.3 | 30 |
| *C. elegans* | 99 | 5,628 | 17,585 | 0.25 | 469 | 49.5 | 75 |
| *D. melanogaster* 180 Mbp | ~60* 120 | $\delta$ ~ 13,235 $\delta$ ~ 8,823 | 13,600 | 0.13 | 573 | 46.8 | 87 |
| Human (old) ~3,000 Mbp | 1,000* 2,000 | $\delta$ ~ 30,000 $\delta$ ~ 20,000 | 100,000? | ~0.03 | ~300? | ~180.0? | ~97? |
| Human (update) ~2,900 Mbp | 967* 1,933 | $\delta$ ~112,500 $\delta$ ~75,000 | ~25,800 | ~0.016 | ~600 | ~92.9 | ~98.4 |

- highly repeated sequences.

## 3 Figures of the Human Genome

There had been many attempts to estimate the number of human genes indirectly. In the mid-1980s, it was suggested that there might be about 100,000. This estimate, widespread in 80's and late 90's, was based on a typical genome of ~3 Gbp (bp=base pairs) after eliminating highly repetitive sequences and assuming a human coding density in the order of 20,000 bp/gene [17]. This figure was led to being widely quoted in many textbooks [15-17]. Until the end of the previous century, most guesstimates of the number of genes in human beings ranged from 50,000 to 100,000. Estimates based on ESTs suggested 120,000 genes [18]. Extrapolating from the number of CpG islands with known genes made an estimate of 70,00-80,000 human genes [19]. The analysis of sequence tags indicated 35,000 genes [20]. Comparison of whole-genome shotgun sequence from the pufferfish with the human genome was used to estimate the density of 'exons', suggesting around 30,000 human genes [21].

This paper corroborates the paucity of human genes currently accepted [22-23]. Our claim is substantiated by taking into account the average protein length expected for humans. Deloucas et al., 1998 [24] proposed a physical map of 30 kgenes, but rooted in customary estimates of that epoch, they argued, "…containing perhaps half of all human genes" (*sic*). This rough calculation of the number of human genes shows no discrepancy with further estimates derived since the human genome sequence was published [25].

The estimated number for the number of protein-coding genes of higher eukaryotic organisms is usually somewhat different, due to their particular DNA structure. Brief comments on previous estimates of the number of human genes are also presented. Let $C'$ and $\delta'$ denote, respectively, the genome size and the coding density with the exception of highly repetitive sequences [26]. About one third of high eukaryotic DNA corresponds to these sequences, which are not transcribed, but may have structural properties [17]. Therefore, $C'=2C/3$ and $\delta'=2\delta/3$. The number of genes can be estimated according to the formula

$$g = \frac{C}{\delta} = \frac{C'}{\delta'} \quad \text{genes.} \qquad (4)$$

The superscript "prime" refers to the expurgated genome, i.e. highly repeated sequences apart.

The largely widespread estimate until late 90's for human genome assumed $C' \approx 2,000$ Mbp and $\delta \approx 20,000$ bp/gene in Eqn(4), thereby yielding 100,000 genes [17]. However, no account was taken of the fact that such a density leads to an average polypeptide barely 300 amino acid residue in length. Human beings may have $\overline{L} \approx 600$ aa, so a much more realistic estimate from Eqn(2) gives $g \approx 48,300$ genes. Nevertheless, the key most up-to-date refinement must be concerned with the genomic rate, assumed to be $R=1.6\%$ instead of 3.0% [22] yielding $g \approx 25,800$. Values for the coding density $\delta'$ can be estimated now from $\delta' = C'/g$ (Table 2).

Many times, it was not clear whether pseudogenes were expurgated or not in the number of genes guesstimates. This fact partially accounts for some misunderstanding on the gene amount of the human genome.

Table 3 presents an expected gene distribution in the 23 human chromosomes, considering the computerized DNA database[1]. This rough calculation of the number of human genes shows no discrepancy with further estimates derived since the human genome sequence was published [25].

Table 3. A plausible gene distribution in the 23 human chromosomes: Genome size $C$=2,881 Gbp; Number of genes $g$=22,525. Despite the fact that this distribution is rather speculative, it may furnish a guideline on what number of genes is to be expected in a particular chromosome. The unveil number of genes in the last column is extracted from the URL http://www.nature.com/nature/focus/humangenome/5.html

| chromosome | length (bp) | predicted genes (unveiled genes) |
|---|---|---|
| Chr1 | 226,828,929 | 2,016 |
| Chr2 | 205,000,000 | 1,822 (1,346) |
| Chr3 | 195,073,306 | 1,734 |
| Chr4 | 115,000,000 | 1,022 (796) |
| Chr5 | 117,696,509 | 1,046 (923) |
| Chr6 | 169,212,327 | 1,504 (1,557) |
| Chr7 | 310,210,944 | 1,367[a] (1,150) |
| Chr8 | 143,297,300 | 1,274 |
| Chr9 | 117,790,386 | 1,047 (1,149) |
| Chr10 | 132,016,990 | 1,173 (816) |
| Chr11 | 130,908,954 | 1,163 |
| Chr12 | 129,826,379 | 1,154 |
| Chr13 | 90,000,000 | 800 (633) |
| Chr14 | 87,191,216 | 775 (1,050) |
| Chr15 | 81,992,482 | 729 |
| Chr16 | 79,932,432 | 711 (880) |
| Chr17 | 79,376,966 | 705 |
| Chr18 | 74,658,403 | 663 |
| Chr19 | 55,878,340 | 497[b] (1,461) |
| Chr20 | 59,424,990 | 528 (727) |
| Chr21 | 33,924,367 | 301[c] (225) |
| Chr22 | 34,352,072 | 305 (545) |
| Chr X | 152,118,949 | 1,352 (1,098) |

a adjusted value to a chromosome length of 153,800,000
b the chromosome 19 is known to hold the highest (unusual) gene density of all human chromosomes
c the chromosome 21 is recognized as an exceptionally gene-poor chromosome.

Many human chromosomes have already been examined and their genes identified. The agreement between these findings and the gene prediction presented in this paper can be checked (Table 3).

This plausible gene distribution in the 23 human chromosomes is obviously a mere guideline for the expected number of genes. Although every one of the approaches for approximately calculating the quantity of genes is just accurate enough to provide an order of magnitude, the reasoning presented here — even if somewhat speculative — is an additional sign that fewer genes than 25 kgenes are to be anticipated. The ultimate answer is expected to be discovered shortly.

---

[1] http://www.cbs.dtu.dk/services/GenomeAtlas/show-genus.php?kingdom=Eukaryotes&genus=Homo&species=sapiens&strain=Strain

The genes mean size $\overline{gene}$ (bp) in each chromosome is given by

$$\overline{gene} = e.\overline{E} + (e-1).\overline{I} . \qquad (5)$$

Characteristics of genes into a few human chromosomes are compiled in Table 4. The values of the chromosome size, the mean 'exon' length, the mean 'exons' per gene and the gene mean size were collected from the references for chromosomes 6, 10, 13, 20 and 22. For Chr9 and 14, the mean 'exons' per gene was derived by dividing the total number of genes by the number of genes of the respective chromosomes. Values of the mean 'intron' length $\overline{I}$ were derived from eqn(5).

Table 4. Identified genes into some human chromosomes (Chrom.) For each chromosome, the mean 'exon' length ($\overline{E}$), the mean 'intron' length ($\overline{I}$), the mean 'exons' per gene (*e*) and genes mean size ($\overline{gene}$) are also shown.

| Chrom. number | C (bp) | genes & pseudo (only genes) | $\overline{E}$ (bp) | $\overline{I}$ (bp) | e | $\overline{gene}$ (kbp) |
|---|---|---|---|---|---|---|
| Chr2 [27] | 237,000,000 | 2,585 (1,346) | -- | -- | 5.30 | 33.8 |
| Chr4 [27] | 186,000,000 | 1,574 (796) | -- | -- | 6.60 | 34.3 |
| Chr6 [28] | 166,800,000 | 2,190 (1,557) | 318 | 7,208 | 5.28 | 32.5 |
| Chr9 [29] | 109,044,351 | 1,575 (1,149) | 342 | 6,799 | 5.77[a] | 34.4 |
| Chr10 [30] | 131,666,441 | 1,357 (816) | 322 | 7,817 | 5.84 | 39.7 |
| Chr13 [31] | 95,500,000 | 929 (633) | 320 | 9,164 | 5.20 | 40.2 |
| Chr14 [32] | 87,410,661 | 1,443 (1,050) | 295 | 8,194 | 6.35[a] | 45.7 |
| Chr20 [33] | 59,187,298 | 895 (727) | 292 | 5,170 | 6.00 | 27.2 |
| Chr22 [34] | 34,491,000 | 679 (545) | 266 | 4,037 | 5.40 | 19.2 |

[a] obtained from: No. of exons/ No. of genes.

The average number of amino acid residues ($\overline{L}$), derived by combining Eqns (1) and (3), is shown in Table 5 for each chromosome, corroborating our initial guess. The genomic rate of a specific chromosome can be obtained from

$$R = e\overline{E}g/C . \qquad (6)$$

Table 5. Identified human genes. For each chromosome, the average number of amino acid residues ($\overline{L}$) and the genomic rate (*R*) are shown.

| Chrom. number | Chr6 | Chr9 | Chr10 | Chr13 | Chr14 | Chr20 | Chr22 |
|---|---|---|---|---|---|---|---|
| $\overline{L}$ (aa) | 560 | 658 | 627 | 555 | 624 | 584 | 479 |
| R (%) | 1.56 | 1.79 | 1.17 | 1.10 | 2.36 | 2.15 | 1.82 |

## 4 Conclusion

This short note discussed about genome figures with focus on its average length of polypeptide chains of proteins expressed by the genes. Relationships were derived among parameters such as genome size, genomic rate, coding density, average number of aa residues of proteins, average length of 'exons' and average number of 'exons per gene'. The human genome was specially considered, presenting a plausible gene distribution in human chromosomes. It was shown that it presents typically an average length of 'exon' about 300 bp, the average length of 'intron' about 6,900 bp, there is a mean of about 6 exons/gene (from single-exon genes to 175 exon for the Titin gene!) and the average number of residues for coded-proteins is close to 600 aa. Finally, the preliminary numbers of this study also points out to the average protein size as a worthy criterion for assessing life complexity.

This study was partially supported by the Brazilian National Research Council (CNPq) through research grants # 306049 (NSSM) and # 306180 (HMdO).